\documentstyle[amssymb,graphicx,twoside,fleqn,espcrc2]{article}

\newcommand{\fig}[1]{{\frenchspacing Fig.~(\ref{#1})}}

\newcommand{\beq}{\begin{equation}}
\newcommand{\eeq}{\end{equation}}
\newcommand{\figs}[1]{\centering{{\includegraphics[width=7cm,height=4.15cm]{FIGS/#1}}}}
\newcommand{\figsm}[1]{\centering{{\includegraphics[width=3.8cm]{FIGS/#1}}}}
\newcommand{\figsmm}[1]{\centering{{\includegraphics[width=7cm]{FIGS/#1}}}}
\newcommand{\figsmmm}[1]{\centering{{\includegraphics[width=3.5cm]{FIGS/#1}}}}
\newcommand{\figsmml}[1]{\centering{{\includegraphics[width=5.3cm]{FIGS/#1}}}}
 
\newcommand{\C}{\mathbb{C}}

\title{Efficient computation of low-lying eigenmodes of non-Hermitian
Wilson-Dirac type matrices}

\author{H.\ Neff \address{John von Neumann Institute for Computing
Research Center J\"ulich, 52425 J\"ulich, Germany\\
neff@theorie.physik.uni-wuppertal.de}}

\begin{document}

\begin{abstract}
  A polynomial transformation for non-Hermitian matrices is presented,
  which provides access to wedge-shaped spectral windows.  For
  Wilson-Dirac type matrices this procedure not only allows the
  determination of the physically interesting low-lying eigenmodes but
  also provides a substantial acceleration of the eigenmode algorithm
  employed.
\end{abstract}

\maketitle

\section{Introduction}

The spectra of Wilson-Dirac type matrices $M$ have an elliptic shape,
with the real parts of the eigenvalues being positive. Since the large
eigenmodes correspond to the doublers, only the small ones should
contain physics.

Employing the Arnoldi algorithm \cite{PARPACK}, the straightforward
approach to the described eigenproblem is to ask the algorithm for the
eigenmodes of smallest modulus. By introducing shifts along the real
axis ($M+\sigma I$), the part of the spectrum to be found by the
algorithm can be influenced.

On the other hand we observe that the Arnoldi algorithm converges much
faster when asking for eigenmodes of largest modulus or largest real
part of $(-M + \sigma I)$ instead, where on a quenched $4^4$ ($8^4$)
lattice a gain factor of about 2 (5) in CPU time can be achieved. The
disadvantage of this approach is that the computed eigenmodes are not
as 'low-lying' as before (in general their imaginary parts are
larger). This problem can be cured by a modified approach, which will
be described in the following.

\section{The polynomial transformation $p_{\sigma, n}$}

In order to focus the search for large eigenmodes of $(-M +
\sigma I)$ on the ones close to the real axis (i.e.\ on low-lying
eigenmodes of $M$), we propose to determine 
the eigenmodes of {\it largest real parts of}
\begin{equation} \label{EQ:poly}
p_{\sigma, n} (D) = \left( D + \sigma I \right)^{n} \; ,
\end{equation}
with $D$ defined through $M=I -\kappa D$. Using $D$ is only a matter
of notation, one could as well work with $p(M)=( -M + \sigma I
)^{n}$.\footnote{For polynomial acceleration of eigenmode
algorithms, see e.g.\ \cite{SAAD}}

For the standard Wilson-Dirac matrix, $D$ can be even-odd
preconditioned.  The polynomial transformation then takes the form
\beq \label{EQ:poly_eo} p_{\sigma, n}(D_{eo} D_{oe} ) =
\left(D_{eo} D_{oe} + \sigma I \right)^{n}\; , \eeq where $e$
and $o$ stand for even and odd respectively.

\begin{figure}[t] 
\centering{\mbox{ 
\figsmmm{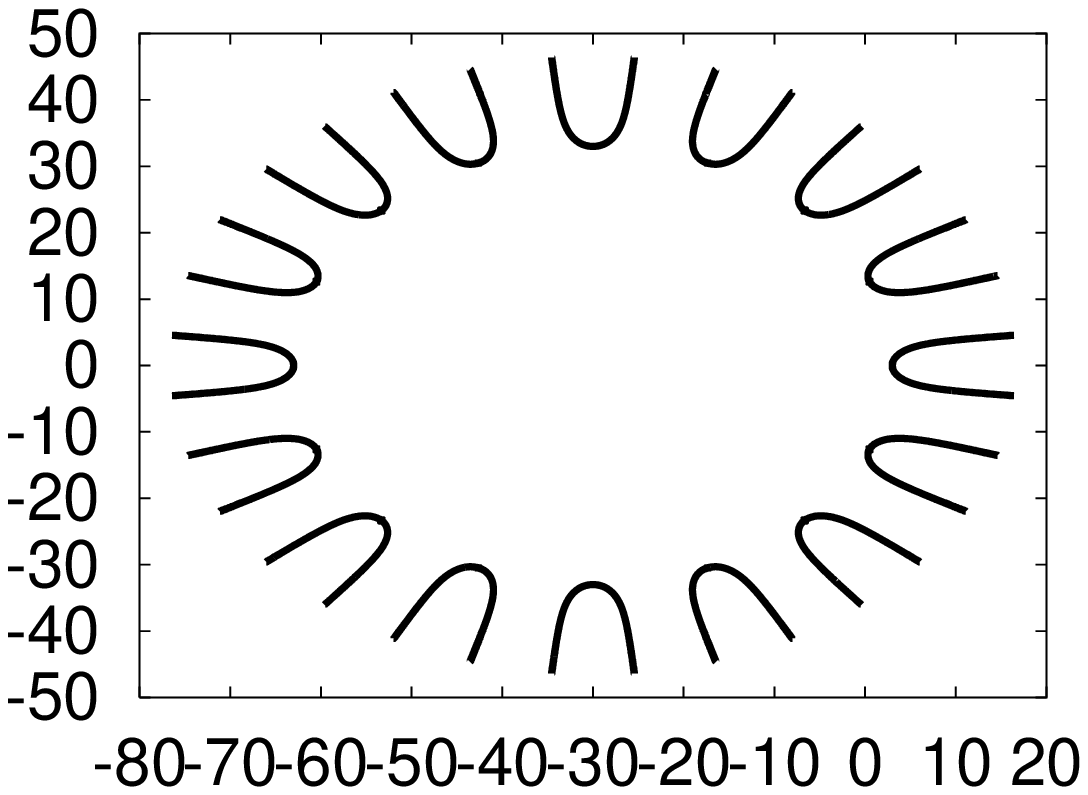} \figsmmm{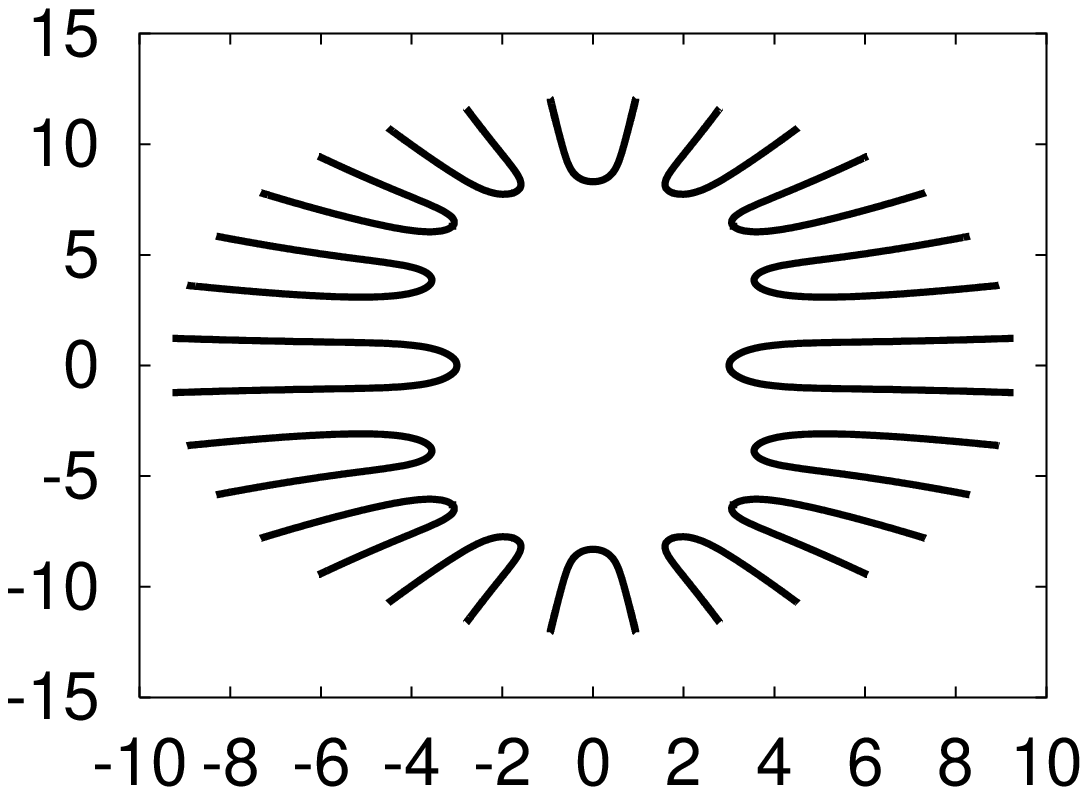}}}
\vskip-0.7cm \caption{Spectral windows of sensitivity. Left: without
preconditioning, for $n=16$, $\sigma=30$ and
$c=(\sigma+3)^{n}$, right: with even-odd preconditioning,
for $n=8$, $\sigma=30$ and $c=(\sigma+3)^{n}$.}
\label{FIG:reg_sens}
\vskip-0.5cm
\end{figure}

The spectral windows amenable to $p_{\sigma, n}$ can be identified by
determining those complex numbers, $a$, which are mapped into $ S= \{
a \in {\C} \; | \; \mbox{real} ( p_{ \sigma, n }(a)) \ge c \}$, where
$c$ depends on the number of eigenmodes to be determined.  This is
illustrated in \fig{FIG:reg_sens}. The curves enclose the complex
numbers belonging to $S$, with their shape being fixed by $\sigma$ and
$n$.  With respect to the eigenmode algorithm, $S$ represents the
regions of sensitivity, i.e.\ the algorithm is capable to find
eigenmodes lying inside the wedge-like regions $S$.

\begin{figure}[t]
\centering{\mbox{ \figsmml{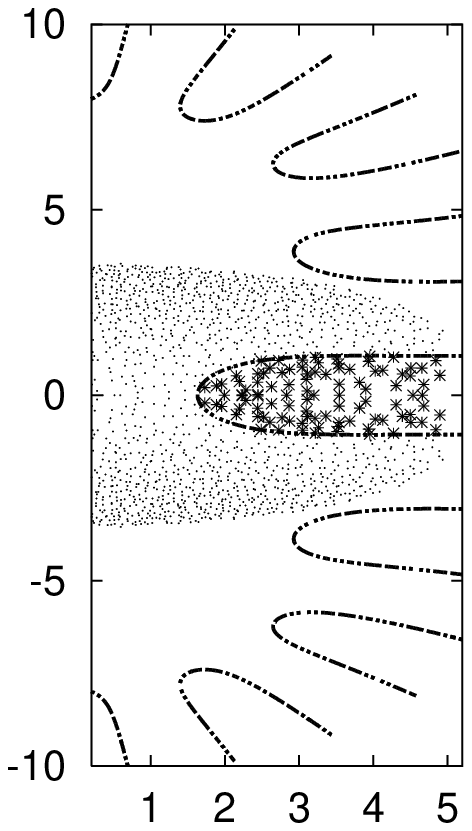} \hskip-3.1cm
\figsmml{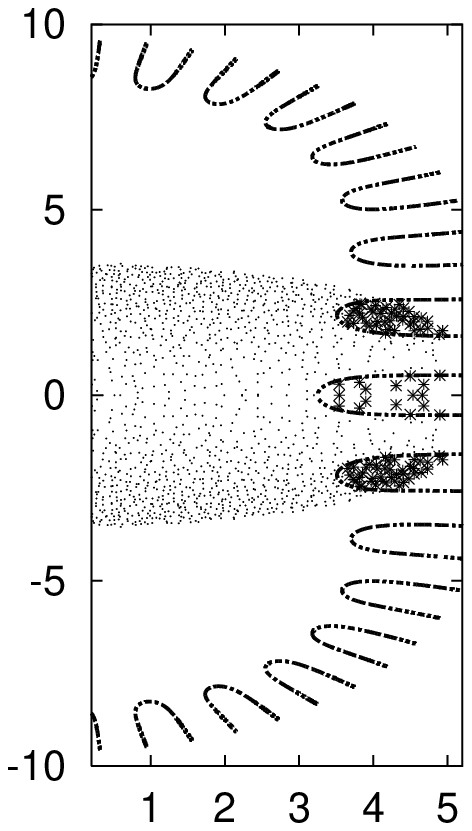}\hskip-3.1cm \figsmml{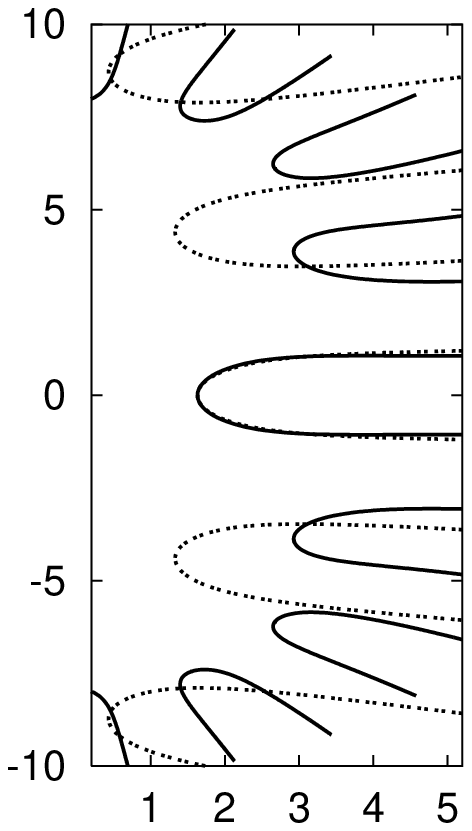}}}
\vskip-0.7cm \caption{100 eigenvalues (fat points), determined on a
quenched $4^4$ lattice for $\beta=5.0$. Left: $p_{\sigma,n}$
with $\sigma=30$ and $n=8$, even-odd preconditioned. Middle:
$p_{\sigma,n}$ with $\sigma=30$ and $n=16$, even-odd
preconditioned.  Right: Comparison of $p_{30,46}$ without even-odd
preconditioning (dotted lines) with $p_{30,8}$ for even-odd
preconditioned matrix (full lines).}
\label{FIG:ev_eo_reg_sens}
\vskip0.3cm 
\centering{\mbox{
\figsmmm{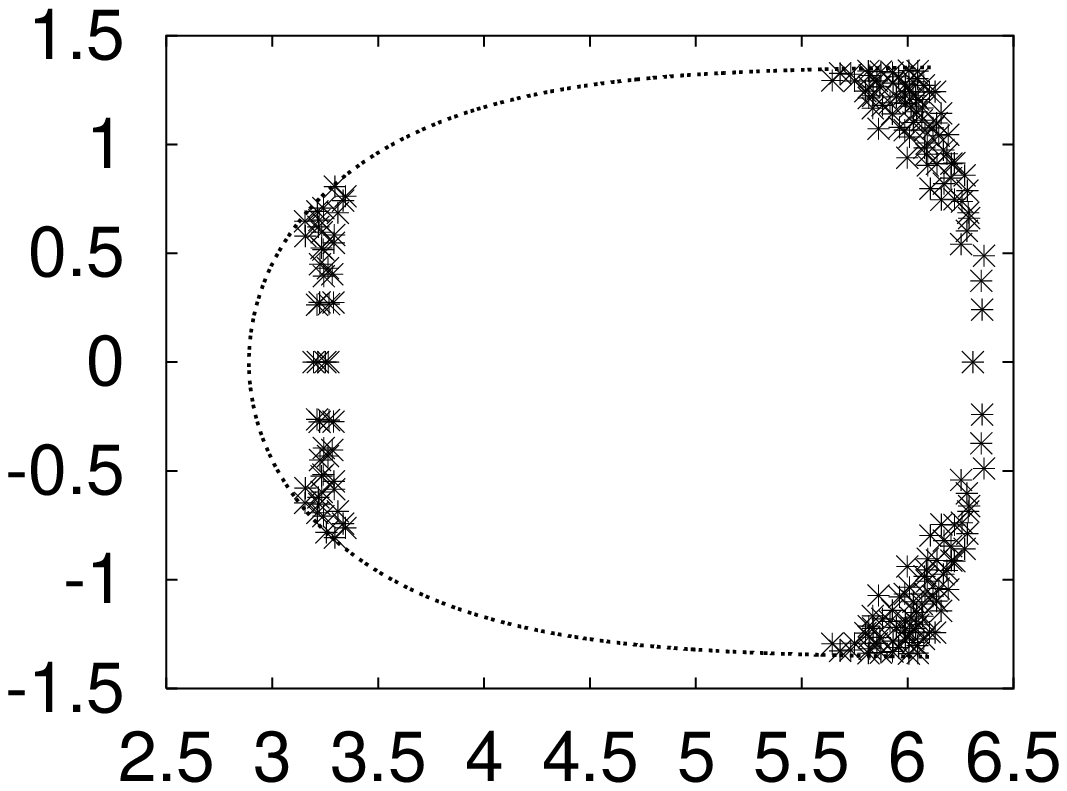} \figsmmm{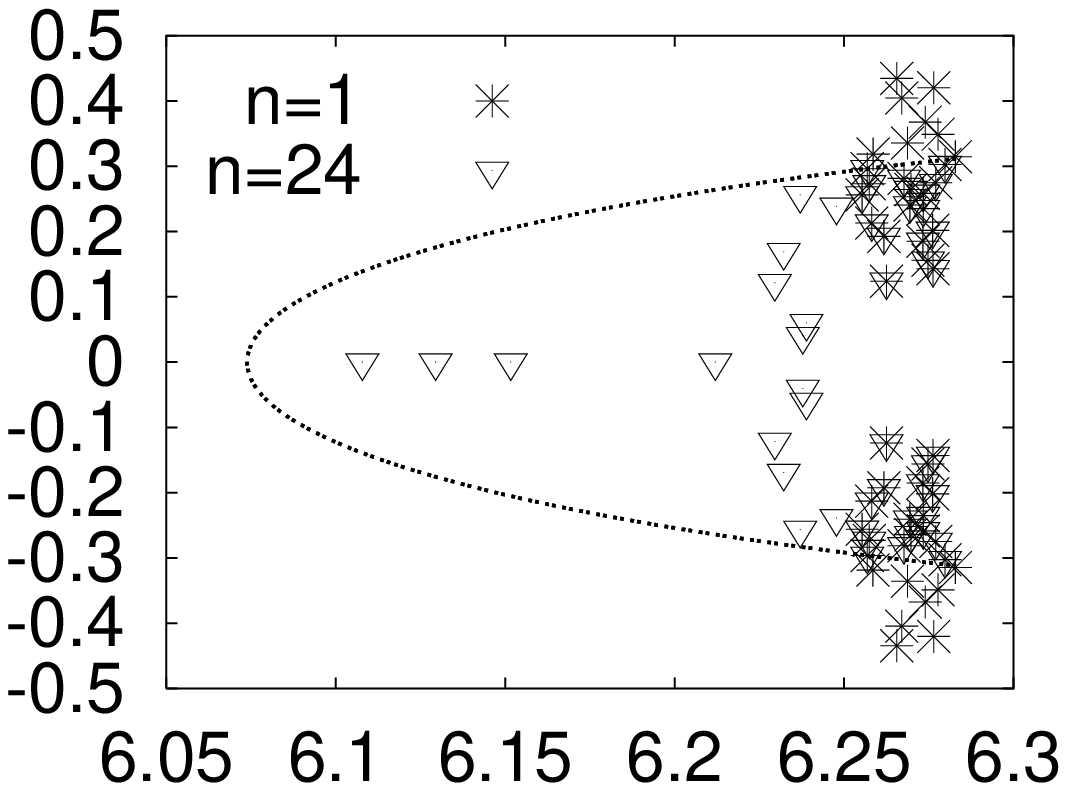}}}
\vskip-0.7cm \caption{Left: 200 eigenvalues on a quenched $8^4$
lattice, for $\beta=6.0$, $\sigma=50$ and $n=8$.  Right: 50
eigenvalues on a full QCD $16^3 \times 32$ lattice, for $\beta=5.6$
and $\sigma=50$. The curve corresponds to $n=24$.}
\label{FIG:ev_examples}
\vskip-0.5cm
\end{figure}

\begin{figure}[h]
\centering{\figs{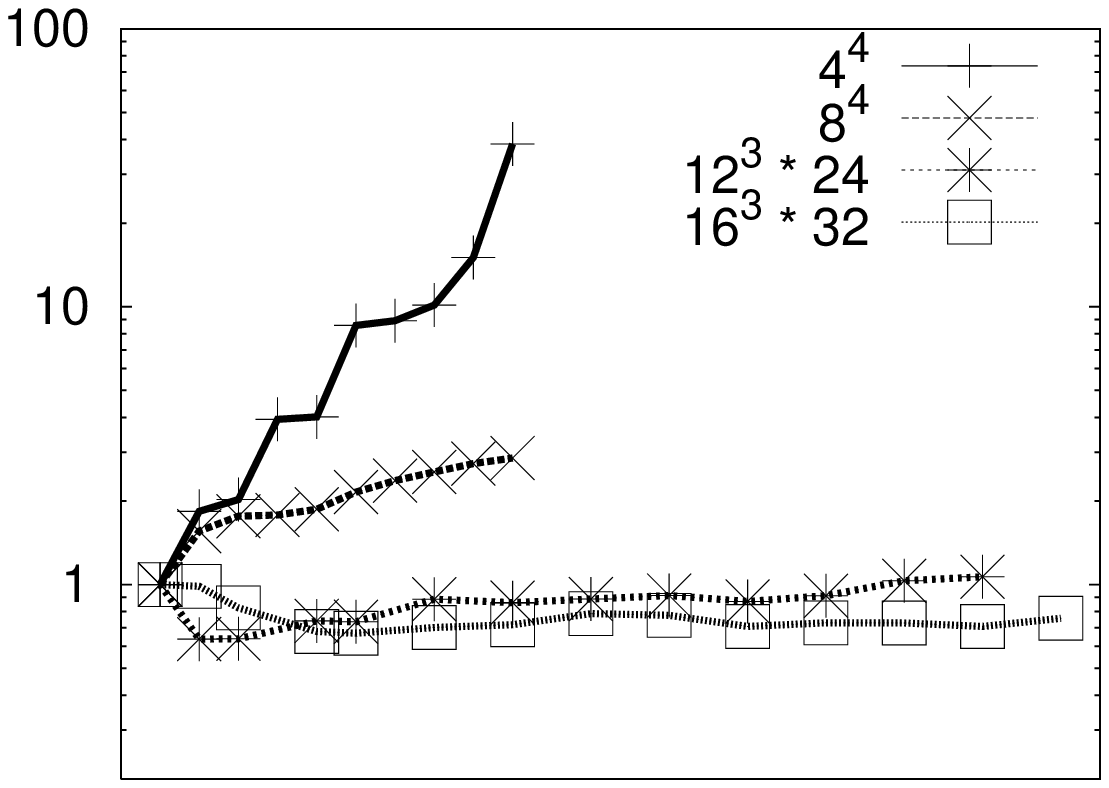}}
\vskip-0.6cm
\centering{\figsmm{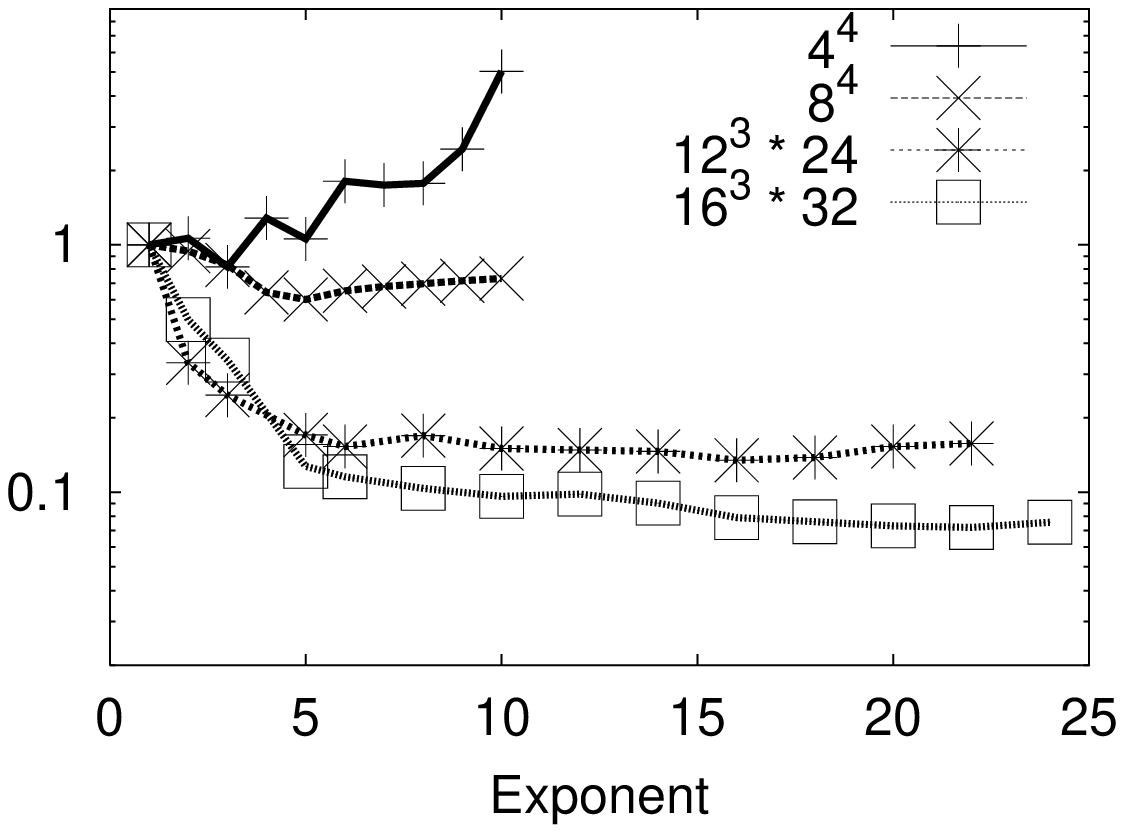}}
\vskip-0.7cm \caption{Dependence of the number of matrix vector
multiplications (upper frame) and the CPU time (lower frame) on the
exponent $n$ (normalized such that for $n=1$ all values
are equal to 1). 50 eigenmodes of the even-odd preconditioned
Wilson-Dirac matrix were determined with $\sigma=50$.  The $4^4$,
$8^4$ and $12^3 \times 24$ lattices are quenched.}
\label{FIG:runtime_M_eo}
\vskip-0.7cm
\end{figure}

\begin{figure}[t]
\centering{\mbox{
\figsm{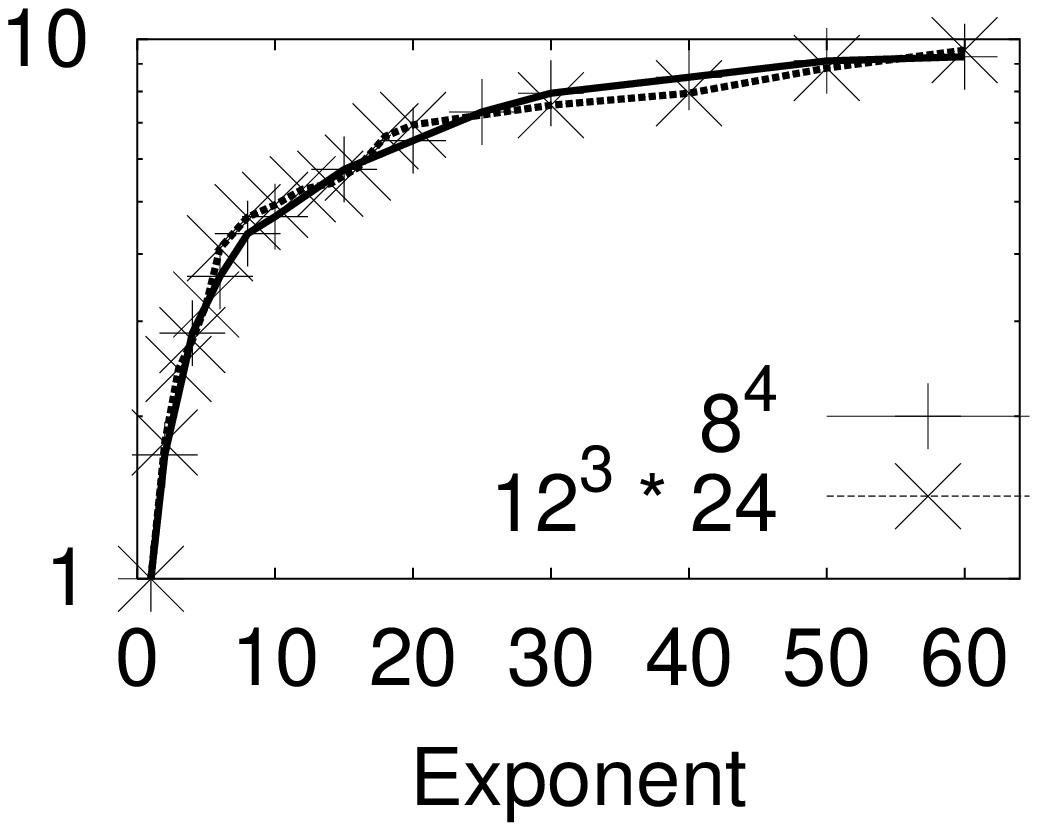} \hskip-0.5cm \figsm{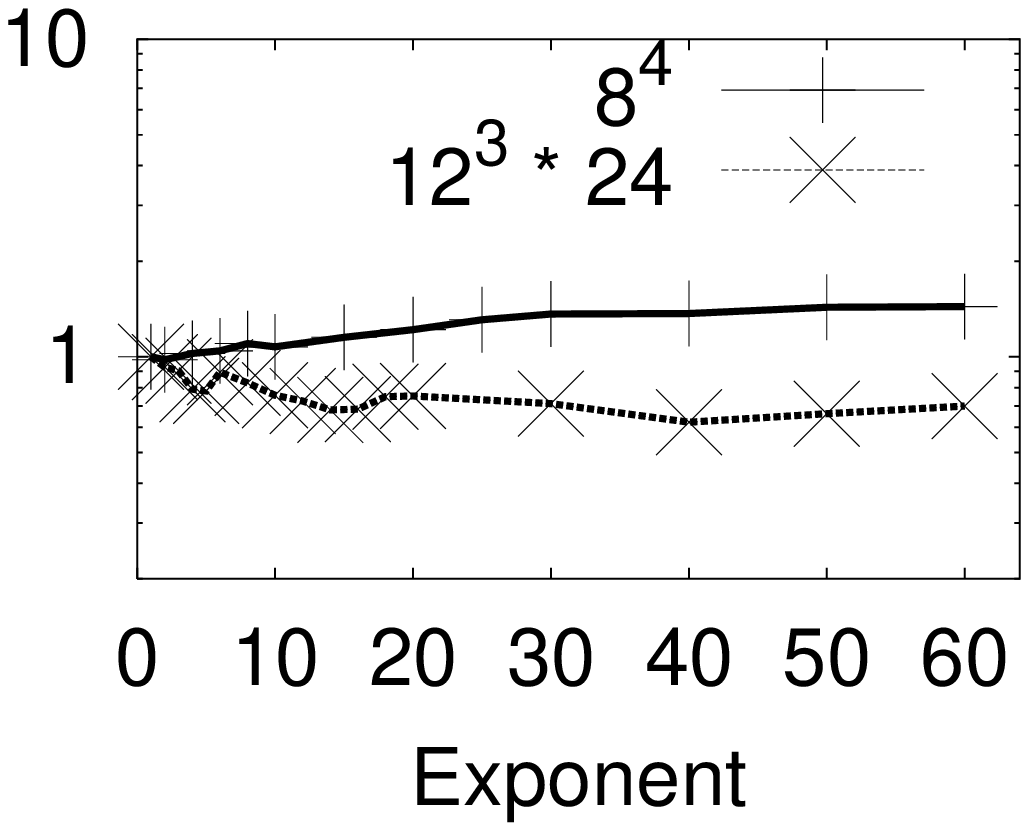}}}
\vskip-0.6cm \caption{Dependence of the number of matrix vector
multiplications (left) and the CPU time (right) on the exponent
$n$ (normalized such that for $n=1$ all values are equal
to 1). 50 eigenmodes of the clover improved Wilson-Dirac matrix were
determined with $\sigma=50$. Both lattices are quenched.}
\label{FIG:runtime_M_clover}
\vskip-0.2cm
\end{figure}

\section{Results}

The spectral windows of the standard and the clover improved
Wilson-Dirac matrix amenable to $p_{\sigma,n}$ are displayed in
\fig{FIG:ev_eo_reg_sens}.

The left frame shows  that the computed eigenmodes
are indeed enclosed by the curves defined through $S$ and that they
are close to the real axis, i.e.\ low-lying.

Furthermore the second frame illustrates that with too large an
exponent $n$, physically uninteresting eigenmodes with large
imaginary parts will be computed by the algorithm.

Finally the right frame demonstrates that preconditioning can
help to substantially reduce $n$ without changing the interesting window of
sensitivity, as the curves cutting the real axis nearly coincide.

In \fig{FIG:ev_examples} further such examples are given. The left
frame shows 200 eigenmodes as found on a quenched $8^4$ lattice. The
right frame compares, for a full QCD $16^3 \times 32$ lattice, the
eigenmodes found for the exponents $n=1$ and $n=24$.
The plotted curves enclose the regions of the complex plane that were
searched for eigenvalues, i.e.\ inside the curves all eigenvalues
are captured.

One might suspect that the polynomial transformation
$p_{\sigma,n}(D)$ goes along with a polynomial increase of the
execution time due to additional matrix vector multiplications.  That
this is not so can be seen for the even-odd preconditioned matrix in
\fig{FIG:runtime_M_eo} and for clover improvement in
\fig{FIG:runtime_M_clover}, where the dependence of the number of
matrix vector multiplications and of the CPU time on the exponent
$n$ are displayed.\footnote{Surprisingly in some cases only for
$n=4$ no convergence was found.}

For the even-odd preconditioned Wilson-Dirac matrix one can see that
the run time increases strongly on the $4^4$ and is almost stable on
the $8^4$ lattice, whereas a dramatic decrease is found on the
realistic lattice sizes $12^3 \times 24$ and $16^3 \times 32$.  For
the last two lattices the CRAY T3E-1200 found an acceleration factor
of 8, whereas the CRAY T3E-600 produced a gain factor of 9 and 14
respectively.

For the even-odd preconditioned Wilson-Dirac matrix this systematics
seems to be delayed. On the $8^4$ lattice an increase in computing
time is detected, whereas on the $12^3 \times 24$ an acceleration
factor of $1.6$ was found.

Furthermore the results indicate that the factor of acceleration grows
with the size of the lattice.

{\it Why does the polynomial transformation accelerate the algorithm?}

At first one might expect that through the polynomial transformation
the density of the eigenvalues is decreased. This would explain the
acceleration, since the convergence is sensitive to this quantity. But
if that is the right explanation, the number of matrix vector
multiplications would be decreasing as well. Since this is not always
the case (see \fig{FIG:runtime_M_eo} and \fig{FIG:runtime_M_clover}),
a different mechanism must be at work.  Taking a look at the time
spent in the Arnoldi subroutines for different exponents $n$
immediately provides the answer: When employing the polynomial
$p_{\sigma,n}$ one has to enter the Arnoldi subroutines less often,
since in one Arnoldi step one performs $n$ multiplications instead of
only one and thus saves the time normally spent in those routines.
This means that for large $n$, the computing time is almost
exclusively used for the matrix vector multiplications. Hence the
optimal choice of polynomial is met, if just one Arnoldi factorization
has to be performed.


\section{Summary and comments}

It could be shown that through the polynomial transformation
$p_{\sigma, n}$ wedge-shaped spectral windows are
accessible. In case of the even-odd preconditioned Wilson-Dirac matrix
this led to an acceleration factor of an order of magnitude on
typical lattices sizes, whereas for the clover improved Wilson-Dirac
matrix a factor of the order of 2 was found. It could be shown that
this factor grows with the size of the lattice. Comparing with a
search for the eigenmodes of smallest modulus of $M$, which provides a
similar part of the spectrum as $p_{\sigma, n}(D)$, the factor
of acceleration is even larger.

The possible accuracy with which one can determine the eigenmodes is
not affected by the polynomial transformation.

{\bf Acknowledgments} I thank Ivan Hip for his help and the staff of
the NIC Research Center J\"ulich, where the computations were carried
out.

\bibliographystyle{h-elsevier}
\bibliography{lit}

\end{document}